\documentclass{appolb}
\usepackage{graphicx}
\newcommand\ltdash{\raise-1.8pt\hbox{$\scriptscriptstyle |$}}
\newcommand \beq  {\begin{equation}}
\newcommand \eeq  {\end{equation}}
\newcommand \bea {\begin{eqnarray} }
\newcommand \eea {\end{eqnarray}}

\newcommand\dg{^{\dagger}}
\newcommand\si{{\sigma}}
\begin{document}
\title{
Hidden Order in $URu_{2 }Si_{2}$: \\ The Need for a Dual Description
\thanks{Presented at the Strongly Correlated Electron Systems 
Conference, Krak{\'o}w 2002}%
}
\author{J.A. Mydosh,$^{1}$ P. Chandra,$^3$ P.Coleman$^4$ and V. Tripathi$^5$
\address{$^1$ Kamerlingh Onnes Laboratory, 
Leiden University,
PO Box 9504, 2300 RA Leiden,
The Netherlands }
\address{$^2$ Max Planck Institute for Chemical Physics of Solids,
01187 Dresden, Germany
}
\address{$^3$ NEC Research, 4 Independence Way,
Princeton, NJ 08540, USA
}
\address{$^4$
Department of Physics and Astronomy,
Rutgers University,
136 Frelinghausen Road,
Piscataway, NJ 08854-8019, USA
}
\address{$^5$Cavendish Laboratory, 
Madingley Road, 
Cambridge CB3 0HE, 
UK
}
}
\maketitle
%
%
%
\begin{abstract}
Motivated by experiment, we argue that
the enigmatic hidden order in $URu_2Si_2$
demands a dual description that encompasses
both its itinerant and its local aspects.
A combination of  symmetry considerations and
observation allow us to rule out a number
of possibilities.  The two remaining scenarios,
the quadrupolar charge density wave and
the orbital antiferromagnet, are discussed
and experiments are suggested to select 
between these proposals.
\end{abstract}
\PACS{78.20.Ls, 47.25.Gz, 76.50+b, 72.15.Gd}
\vskip0.1in

The heavy fermion metal $URu_2Si_2$ undergoes a pronounced
second-order phase transition at $T_0 = 17 K$ characterized
by sharp anomalies in a number of bulk properties
including the specific heat,\cite{Palstra85} 
the linear\cite{Palstra85} and the nonlinear
magnetic susceptibilities,\cite{Miyako91,Ramirez92} where standard mean-field
relations between the measured thermodynamic properties are
satisfied.\cite{Chandra94} 
Neutron scattering measurements\cite{Walter86,Mason91,Broholm91} 
indicate gapped, propagating magnetic excitations that
suggest the formation of a spin density wave.  However the
magnitude of the observed moment\cite{Broholm91} ($m_0 = 0.03 \mu_B$)
cannot account for the entropy loss at the transition,
which has been attributed to the development of an enigmatic hidden order.
Furthermore there is strong evidence from muon spin 
resonance\cite{Luke94} and from pressure-dependent
NMR experiments\cite{Matsuda01} that the magnetic and the hidden ordered
phases are phase separated, and thus develop independently.\cite{Chandra02}
In this paper we review the constraints that recent experimental developments
and symmetry arguments place on the nature of the hidden order in $URu_2Si_2$,
and discuss specific theoretical proposals that emerge consistent
with known measurements.

It is important to distinguish two distinct aspects of the mysterious
phase transition at $T_0$ in $URu_2Si_2$.  The presence
of a Schottky anomaly at $60 K$in the specific heat\cite{Palstra85} 
and the development of a dispersing mode at 
$T \le T_0$ observed by inelastic neutron
scattering\cite{Broholm91} both suggest the importance of
local crystal-field excitations at the transition. 
Nevertheless
a purely local picture {\sl cannot} provide a straightforward
explanation for the observed elastic response\cite{Luthi93} at $T_0$
and yields a field-dependence for the gap\cite{Santini00} that is
{\sl distinct}
from that associated with observed thermodynamic 
quantities.\cite{Mentink96}
Furthermore the clear signatures\cite{Palstra85} of 
Fermi liquid behavior above $T_0$
and the mean-field nature of the transition\cite{Chandra94}
suggest that an itinerant density-wave is involved.
Thus any microscopic model for the hidden order in $URu_2Si_2$
must ultimately reconcile the local electronic physics
of the strongly interacting uranium ions with the fluid
aspects of the heavy-electron phase.

The duality model of Kuramoto and Miyake\cite{Kuramoto90}
is a natural way to treat the localized and itinerant aspects
of $URu_2Si_2$ within a single scheme.  In this approach,
the itinerant excitations of the Fermi liquid are constructed
from the low-lying crystal-field multiplets of the uranium ion.
The quasiparticles associated with the heavy-electron Fermi liquid
in this system are then composite objects formed from the 
localized orbital and spin degrees of freedom of the $U$ ions
and the conduction electron fields.\cite{Kuramoto90}.
The phase transition at $T_0$ is a simple Fermi surface
instability of the composite mobile f-electrons.  This approach
was originally adopted by Okuno and Miyake\cite{Okuno98}
to describe the coexistence of the hidden order with a small
moment in $URu_2Si_2$.  
With the new understanding that the hidden ordered phase
does not contain a staggered magnetization, we revisit
this duality scheme and, guided by recent experiment,
study its implications for the nature of the enigmatic order that
develops at $T = T_0$.

We begin our phenomenological approach to the hidden order in 
$URu_2Si_2$ with a consideration of its
allowed symmetries, where constraints will be imposed by experimental
observation.
More specifically, the mean-field
character of the transition at $T_0$ suggests that the itinerant
nature of the hidden order can be described
by a general density wave whose
form factor will yield clues about the underlying local
excitations involved; we expect it to be incommensurate
due to the fact that the observed entropy loss and the accompanying
gap suggest that it results from a Fermi surface instability.  
We begin by considering a class of density
wave with the most general pairing in the particle-hole channel

\begin{eqnarray}
\langle {c} _{\mathbf{k}+\frac{\mathbf{Q}}{2},\sigma }^{\dagger }{c}
_{\mathbf{k}-\frac{\mathbf{Q}}{2},\sigma '}\rangle  & = & 
A_{\mathbf{k}}^{\sigma \sigma '}(\mathbf{Q}),
\label{pairing1}
\end{eqnarray}
where $\mathbf{Q}$ is the incommensurate ordering wavevector,
$\sigma$ is a spin
index and $A_{\mathbf{k}}^{\sigma \sigma '}(\mathbf{Q})$ defines a
general function of spin and momentum.
A phenomenological Hamiltonian for such an order parameter
would take the form
\begin{equation}\label{}
H = \sum_{\vec{k}\sigma }\epsilon_{\vec{k}}
{{c}} \dg _{\mathbf{k}\si }{{c}}_{\mathbf{k}\sigma } 
+ \sum_{\mathbf{k}}\left[ \Delta({\mathbf{Q}}) 
{{c}} _{\mathbf{k}+\mathbf{Q}/2, \sigma}\dg  f_{\si \si'} (\mathbf{k})
{{c}}_{\mathbf{k}-\mathbf{Q}/2,\sigma } + {\rm  H.c.}\right] + 
{\cal  N} \frac{\vert \Delta(\mathbf{Q}) \vert
^{2}}{g}\end{equation}
where $\cal  N$ is the number of sites in the lattice and $f_{\si
\si'}$
is a form factor associated with the incommensurate spin density wave; a
mean-field
treatment yields the expression
\begin{equation}
-g A_{\mathbf{k}}^{\sigma \sigma '}({\mathbf{Q}}) = 
\Delta ({\mathbf{Q}}) f_{\si \si'} (\mathbf{k})
\label{gap}
\end{equation}
Of course, a true
microscopic approach must account for how such a coupling
emerges from the residual interactions amongst the heavy
electrons. 

\begin{table}[h]

\begin{tabular}{|c|c|c|c|}
\hline 
&&&\\
Name&
 $f_{\sigma \sigma '}(\mathbf{k})/\delta_{\sigma \sigma'}$&
 T-reversal&
 Local \\
&&-invariance&Fields\\
\hline
&&&\\
SDW 
&
 ${\sigma}
$&
 no&
 yes\\
(isotropic SDW)&&&\\
\hline
&&&\\
CDW 
&
 const.&
 yes&
 no\\
(isotropic CDW)&&&\\
\hline
&&&\\
d-SDW&
 ${\sigma}
 \, (\cos (k_{x}a)-\cos (k_{y}a))$&
 no&
 no\\
&&&\\
\hline
&&&\\
q-CDW 
&
 $\cos (k_{x}a)-\cos (k_{y}a)$&
 yes&
 no\\
(quadrupolar CDW)
&&&\\
\hline
&&&\\
OAFM 
&
 $i\, (\sin (k_{x}a)-\sin (k_{y}a))$&
 no&
 yes \\
(orbital antiferromagnet)&&&\\
\hline
\end{tabular}
\vskip0.1in
\caption{Possible symmetries for particle-hole pairing}

\label{symmetries}

\end{table}

We now follow the approach of Halperin and Rice,\cite{Halperin68} categorizing
the possible particle-hole pairings in $URu_2Si_2$.
Assuming the the hidden order develops betoween $U$ atoms in each
basal plane, we restrict our attention to nearest-neighbor
pairings on a two-dimensional square lattice\cite{Chandra02} and
display the five resulting possibilities 
in Table I.  
All of these pairings lead to a gapping of the Fermi Surface,
accounting for the large entropy loss and the observed anomalies
in various bulk properties. 
For example, let 
$\Delta \chi $ be the reduction in the Pauli susceptibility
due to this gap, then we can describe the hidden order phase
by a Landau-Ginzburg free energy 
\begin{equation}
F =  a (T-T_{c})\Psi^{2} + b \Psi^{4}  +\frac{1}{2}\Delta \chi B^{2}
\Psi^{2}.
\end{equation}
such that the last term is responsible for the strong field-dependence
of the transition temperature ($T_{c} (B) =  T_{c} +\frac{1}{2a}\Delta
\chi B^{2}$) in this material;\cite{Jaime02}
it is also responsible to why $\chi_3$ has the
same temperature-dependence as 
the specific heat.\cite{Ramirez92,Chandra94}
Unfortunately this aspect
of the problem does not discriminate between the possible density wave
order parameters, and we are forced to make more microscopic
considerations. 

Because of large Coulomb repulsion between the quasiparticles
in $URu_2Si_2$, we rule out isotropic pairing in
the charge density wave channel.  Furthermore,
an s-wave charge density wave would result in an accompanying
lattice distortion, but none is observed.\cite{Luthi93}
Similarly neutron scattering does not support the presence
of a spin density wave in the 
hidden ordered phase.\cite{Broholm91,Matsuda01}
Thus, due to the incompressibility of the heavy fermi liquid,
we are left with the three remaining anisotropic pairing states
(see Table I).  We recall that $URu_2Si_2$ undergoes a transition
to a d-wave superconducting state at $T=1.2K$, suggesting
the importance of antiferromagnetic fluctuations in the associated
normal state; by contrast in a d-SDW scenario\cite{Ramirez92} we would
expect ferromagnetic fluctuations to be favored.

\begin{figure}[here]
\begin{center}
\includegraphics[width=0.75\textwidth]{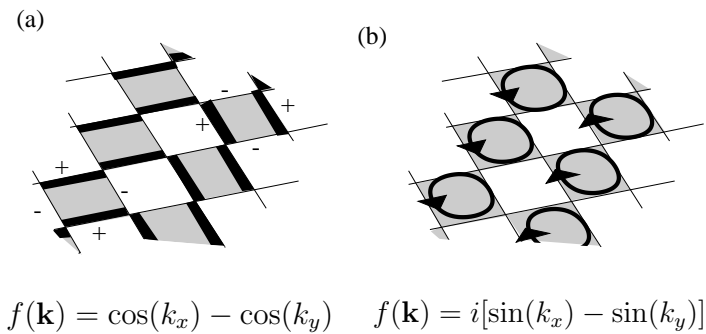}
\end{center}
\caption{(a) Incommensurate quadrupolar density wave (qCDW). In two
dimensions, the form factor $\cos (k_{x})-\cos (k_{y})$ leads to a
an incommensurate density wave with a quadrupolar charge distribution,
the CDW analog of a d-wave superconductor.  (b) 
Incommensurate orbital antiferromagnet.  Here currents circulate
around square plaquettes defined by nearest-neighbor uranium ions.}
\label{fig1}
\end{figure}

This leaves us with two remaining options:  the quadrupolar
charge density wave\cite{Amitsuka02} (Fig. \ref{fig1} (a)) and the orbital 
antiferromagnet (Fig. \ref{fig1} (b)).\cite{Chandra02b}  Both of
these scenarios are consistent with the current picture
of $URu_2Si_2$ as an incompressible Fermi liquid with
strong antiferromagnetic fluctuations. They each have
nodes in the order parameter, so that neither couple directly
to the local charge density.  Furthermore we expect both
incommmensurate density waves to couple weakly to uniform
strain, and thus are both consistent with the observed
insensitivity of the elastic response\cite{Luthi93} at $T_0$.
Recent uniaxial stress measurements suggest
that the hidden order is sensitive to the presence of
local tetragonal symmetry,\cite{Yokoyama02} a feature that is consistent
with both scenarios for completely different reasons.
In the orbital antiferromagnet
the currents are equal in each basal direction,\cite{Chandra02b} whereas
in the quadrupolar charge density wave it is known
that some of the singlet crystal-field states with
tetragonal symmetry are quadrupolar.\cite{Santini00}
We note that the diamagnetic 
\noindent response of the orbital antiferromagnet
is small compared to that associated with the gapping of the
Fermi surface, ($\frac{\chi_{Pauli}}{\chi_{diamagnetic}} \sim
100$),
so that the field-dependence of the hidden order parameter\cite{Jaime02}
cannot be used to discriminate between these two scenarios.

At present, the key factor distinguishing the orbital antiferromagnet
from the quadrupolar charge-density wave scenarios is the presence
or absence of time-reversal breaking.  Orbital antiferromagnetism
is consistent with the isotropy and magnitude of local magnetic
fields\cite{Chandra02b} measured at ambient pressure 
by nuclear magnetic resonance.\cite{Bernal01}  
Such current-carrying states have also been proposed for
one-dimensional
ladders.\cite{Schollwoeck02}  In $URu_2Si_2$  
the line-broadening of the central silicon NMR peak
is the only direct evidence for broken time-reversal symmetry
in the hidden order phase.\cite{Chandra02b}  Recent muon spin measurements 
by Amitsuka and coworkers\cite{Amitsuka02} support the emergence of
local magnetic fields with the temperature dependence
seen in the NMR data,\cite{Bernal01} but the overall amplitude is two
orders of magnitude less than that seen in NMR measurements.
This has led to the proposal that these features may be
an artifact of a minority ferromagnetic phase with dipolar
interactions.  In our opinion, an equally feasible explanation
is that charged muons destroy the hidden order in their
immediate surroundings.  This would explain the similar
but dramatically reduced anomalous magnetic field
measured by the $\mu$SR measurements.  Finally there
have been attempts to test predictions for neutron
scattering but so 
far no clear conclusion has been reached.\cite{Bull02}

There are three types of experiments that would be very
helpful in resolving the nature of the hidden order 
in $URu_2Si_2$:

\begin{itemize}

\item Probes of the nodal quasiparticle structure. 
Both Q-CDW and OAFM have nodes in the
gap.. These nodes would show marked features in the optical
conductivity and the scanning tunnelling spectroscopy; at present
these nodes are inferred but have not been observed directly. 

\item Probes of broken time-reversal symmetry.   Broken time-reversal symmetry
is clearly a critical discriminating feature between the OAFM  and
Q-CDW. The existing NMR measurements strongly support the idea of
broken time reversal symmetry breaking in this material, but further
confirmation is vital.  Further NMR measurements, 
especially on alternative crystal sites, would provide an important way of
confirming whether there are local fields developing at
$T_{o}$.   Careful comparison between the $\mu SR$ and NMR signals
is crucial for deterimining whether 
these two techniques are measuring similar or different phenomena.

\item Revisit old scattering measurements.  There exist
a number of scattering experiments, including polarized neutron and resonant
X-ray, that are suggestive of broken time-reversal symmetry; these
should be redone, given what is now known about 
phase-separation.\cite{Matsuda01}

\end{itemize}

In summary, we believe that a dual description of
the hidden order in $URu_2Si_2$ is necessary 
for capturing its coexisting itinerant and local
features.  The latter is necessary for explaining
the mysterious dispersing mode at $T \le T_0$,
whereas the former is crucial for understanding
the mean-field nature of the transition.  
At present there are two competing scenarios
which differ by a form factors.  We propose
a number of experiments which could resolve
the dispute.

We acknowledge discussions with 
G. Aeppli and H. Amitsuka.
This project is supported
under grant NSF-DMR 9983156 (Coleman and Tripathi).

\end{document}